\documentclass[11pt]{article}

\PassOptionsToPackage{table,dvipsnames}{xcolor}
\usepackage[final]{acl}
\usepackage{times}
\usepackage{latexsym}
\usepackage[T1]{fontenc}
\usepackage[utf8]{inputenc}
\usepackage{microtype}
\usepackage{inconsolata}
\usepackage{amsmath,amssymb}
\usepackage{graphicx}
\usepackage{booktabs,tabularx,multirow,makecell,threeparttable}
\usepackage{array,ragged2e,siunitx}
\usepackage{algorithm}
\usepackage{algpseudocode}
\usepackage{pifont}
\usepackage[normalem]{ulem}
\usepackage{listings}
\usepackage{xurl}
\usepackage{placeins}

\usepackage{CJKutf8}
\usepackage{todonotes}
\usepackage{tcolorbox}
\tcbuselibrary{breakable,skins}
\makeatletter
\renewcommand{\@todonotes@drawInlineNote}{%
  \begin{tcolorbox}[
    enhanced,breakable,
    colback=\@todonotes@currentbackgroundcolor,
    colframe=\@todonotes@currentbordercolor,
    boxrule=0.4pt,arc=2pt,
    left=4pt,right=4pt,top=4pt,bottom=4pt,
    before skip=4pt,after skip=4pt,
    fontupper=\@todonotes@useSizeCommand
  ]%
    \@todonotes@format{\@todonotes@text}%
  \end{tcolorbox}%
}
\makeatother
\newif\ifshowrevisiontodos
\showrevisiontodostrue

\definecolor{armygreen}{RGB}{75,83,32}
\definecolor{deepgreen}{RGB}{0,100,0}
\definecolor{forestgreen}{RGB}{34,139,34}
\definecolor{promptbg}{RGB}{248,248,248}
\definecolor{promptframe}{RGB}{210,210,210}

\lstdefinestyle{promptstyle}{
    basicstyle=\ttfamily\small,
    columns=fullflexible,
    keepspaces=true,
    breaklines=true,
    breakatwhitespace=true,
    showstringspaces=false,
    frame=single,
    framerule=0.4pt,
    rulecolor=\color{promptframe},
    backgroundcolor=\color{promptbg}
}

\title{OmniReview: A Large-Scale Benchmark and LLM-Enhanced Framework for Realistic Reviewer Recommendation}
\author{
  Yehua Huang$^{1}$ \quad 
  Penglei Sun$^{1}$ \quad 
  Zebin Chen$^{1}$ \quad 
  Zhenheng Tang$^{2}$ \quad 
  Xiaowen Chu$^{1,2}$ \\
  $^1$The Hong Kong University of Science and Technology (Guangzhou) \\
  $^2$The Hong Kong University of Science and Technology
}
\hypersetup{pdftitle={OmniReview: A Large-Scale Benchmark and LLM-Enhanced Framework for Realistic Reviewer Recommendation},pdfauthor={}}

\begin{document}
\maketitle

\begin{abstract}
Academic peer review faces a scarcity of large-scale, cross-domain benchmarks and evaluation metrics that reflect editorial workflows.
We present OmniReview, a comprehensive dataset integrating multiple academic data sources through an entity-disambiguation pipeline, with 202,756 verified review records.
We introduce four evaluation tasks covering historical reviewer endorsement, suppression of unsuitable candidates, preference among qualified experts, and identification of qualified but previously unassigned reviewers.
Existing embedding-based approaches suffer from the information bottleneck of semantic compression and limited interpretability.
To address these limitations, we propose Profiling Scholars with Multi-gate Mixture-of-Experts (Pro-MMoE), a framework combining large language models (LLMs) with multi-task learning.
Its LLM-Summary and Raw-text branches capture high-level research intent and preserve fine-grained expertise differences, while a task-adaptive MMoE architecture balances confidence estimation and ranking objectives.
Experiments demonstrate that Pro-MMoE achieves the best overall performance on OmniReview, establishing a benchmark for realistic reviewer recommendation.
Our project is available at \url{https://anonymous.4open.science/r/OmniReview-353D}.
\end{abstract}

\section{Introduction}
\label{sec:intro}

Academic peer review is a standard practice for maintaining scientific quality~\cite{mulligan2013peer}. Given the continuous growth of manuscript submissions, appropriate reviewer assignment is necessary to uphold these standards. 
However, editors often face challenges in identifying experts with the specific domain knowledge and methodological background required for evaluation~\cite{liu2014robust}.
Consequently, reviewer recommendation systems have been developed to support editorial decision-making, using machine learning to match submissions with suitable reviewers~\cite{LIN2023101830,charlin2013toronto,liu2014robust,zhang2025chain}.

High-quality datasets capturing paper--reviewer relationships and expertise underpin both model training and reliable evaluation~\cite{mimno2007expertise,charlin2013toronto,zhang2024oag}.

\begin{figure}[t]
    \centering
    \includegraphics[width=\linewidth]{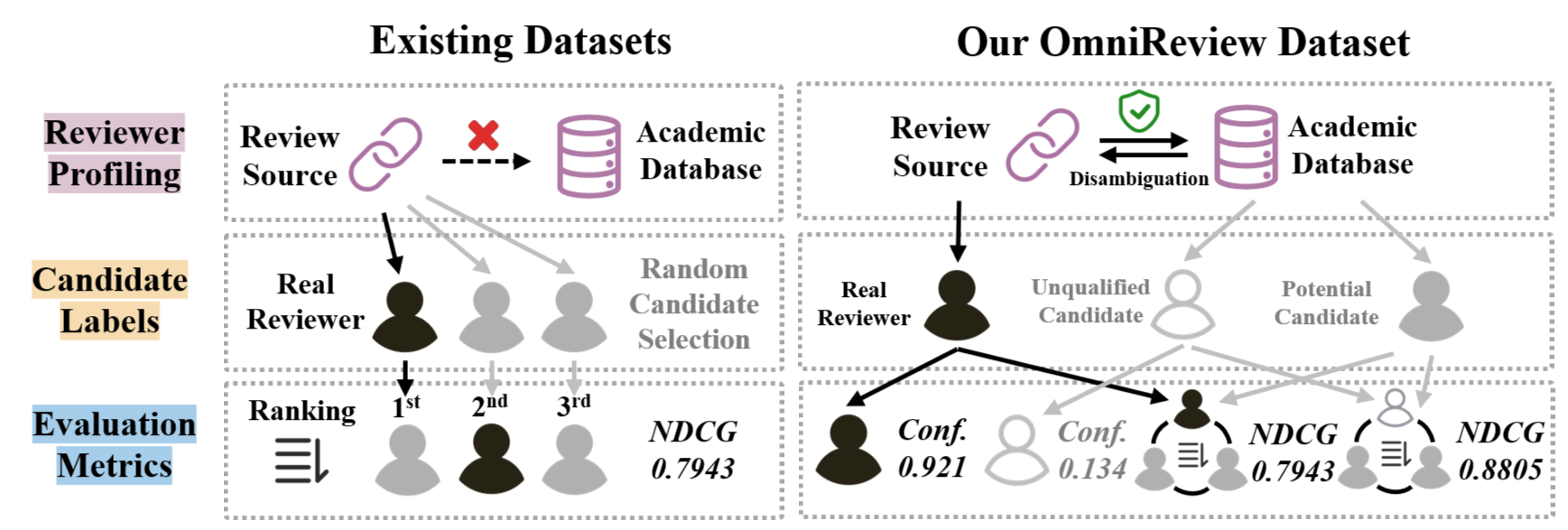}
    \caption{Comparison of existing datasets and our proposed OmniReview dataset.
    Conf. denotes the predicted reviewer suitability score.}
    \label{fig:overview}
\end{figure}

However, existing datasets exhibit three limitations for real-world deployment, as shown in Figure~\ref{fig:overview}: 

\noindent\textbf{(1) Insufficient Reviewer Profiling.} 
Cross-platform scholar identities support research activities~\cite{haak2018using,cress2019academic}, yet existing datasets rely on local metadata and lack comprehensive profiles linked to large-scale academic graphs.
For instance, existing benchmarks~\cite{mimno2007expertise,karimzadehgan2008multi,zhang2025chain} are often confined to basic metadata from specific conferences.
Similarly, other large-scale datasets \cite{peng2025frontier} remain limited to platform-specific information (e.g., within Frontiers), lacking the broader external context required for accurate expertise profiling.

\noindent\textbf{(2) Biased Candidate Labels.} 
Existing candidate construction procedures can introduce two types of bias.
First, datasets constructed using manual annotations or heuristic assumptions may contain label noise. For instance, existing datasets~\cite{mimno2007expertise,zhang2025chain} use researchers' annotations to assess potential reviewers' relevance, which may reflect annotators' subjective judgments.
Additionally, exHarmony~\cite{ebrahimi2025exharmony} treats paper authors as proxy reviewers, although some junior coauthors may not yet possess the qualifications required for reviewing.
Second, relying solely on historical assignment records restricts the candidate pool. 
Some datasets~\cite{peng2025frontier, zhang2024oag} are confined to a closed set of past reviewers. 
However, without explicitly verifying the expertise compatibility between candidates and papers, the recommendation task fails to align with real-world scenarios.


\noindent\textbf{(3) Simplistic Evaluation Metrics.} 
Existing benchmarks~\cite{peng2025frontier,zhang2024oag} primarily focus on standard retrieval metrics based on historical assignments. 
However, given the inherent sparsity of ground-truth labels (i.e., historical reviewers are only a small subset of all scholars), these metrics are affected by false-negative bias.
They penalize qualified but historically unselected candidates and overlook confidence estimation, noise filtering, and discrimination between unassigned experts and unsuitable candidates.

To address these limitations, we present \textbf{OmniReview}, a comprehensive peer-review dataset and evaluation protocol built from review records on the Frontiers open-access platform\footnote{https://www.frontiersin.org/}.
Our approach addresses these limitations in \underline{three ways}.
\textbf{First}, to overcome data fragmentation, we integrate three authoritative sources: the Open Academic Graph (OAG)~\cite{arnetminer,oag1,oag2,sinha2015overview}, the Frontiers open-access platform, and the ORCID Public Data File\footnote{https://info.orcid.org/documentation/integration-guide/working-with-bulk-data/}. 
We use entity alignment, discipline taxonomy construction, and verification to integrate information across these sources.
This process successfully links reviews to researcher profiles, yielding $202,756$ verified records and $150,287$ ground-truth reviewers.
\textbf{Second}, to mitigate label bias and false negatives, we extend the candidate pool beyond historical assignments using publication-supported disciplinary membership.
A hierarchical taxonomy connects each scholar's disciplinary exposure to identifiable publications and provides an explicit basis for selecting potential reviewers, yielding 2,287,697 candidates.
We also examine candidates' h-indices to assess candidate quality and compare their distribution with that of researchers in the academic graph.
\textbf{Third}, we design an evaluation framework with four tasks: (1) Task 1 (Endorsement): Assigning high suitability scores to historical reviewers; (2) Task 2 (Suppression): Assigning low suitability scores to superficially relevant candidates who lack specific domain expertise; (3) Task 3 (Preference): Ranking historical reviewers above other qualified candidates; (4) Task 4 (Qualification): Ranking qualified but previously unassigned reviewers above unqualified candidates.

While embedding-based approaches have shown promise~\citep{Singh2022SciRepEvalAM,zhang2025chain},
they face two challenges in reviewer recommendation.
\textbf{First}, directly embedding raw titles and abstracts is sensitive to wording variation and textual noise,
as similar meanings can produce different representations~\citep{burdick2022paraphrases}.
Incidental phrasing can therefore interfere with matching a target paper's research topics
and methods to a candidate's publications.
\textbf{Second}, embedding scores provide limited interpretability~\citep{zhang2020explainable,zhang2020multi}:
they do not make explicit which publications support a candidate's disciplinary relevance or how those publications align with the target paper.

To address these challenges, we propose \underline{\textbf{Pro}}filing Scholars with \underline{\textbf{M}}ulti-gate \underline{\textbf{M}}ixture-\underline{\textbf{o}}f-\underline{\textbf{E}}xperts (\textbf{Pro-MMoE}),
a dual-path multi-task reviewer reranking framework.
The LLM-Summary Branch distills each publication's research question, methods,
and findings to emphasize core content and reduce phrasing noise.
This summary-based semantic view~\citep{sarthi2024raptor} complements the Raw-text Branch,
which retains original textual details.
Together with OmniReview's discipline taxonomy,
the readable summaries connect candidate selection and research alignment to explicit publication evidence.
Based on this representation, we introduce a task-adaptive MMoE module with shared experts and task-adaptive towers for three training objectives that support four evaluation tasks.
This dual-branch design improves interpretability and preserves textual details, while task-adaptive routing learns objective-specific decision boundaries from shared expert representations.
Experiments on OmniReview demonstrate competitive performance, with the best scores on five of six reported metrics.

In conclusion, this paper makes the following contributions:

\noindent$\bullet$ We construct a large-scale, cross-domain peer-review dataset \textbf{OmniReview},
comprising $202,756$ review records
and $2,287,697$ review candidates involving $150,287$ historical reviewers with manual verification.
Based on this, we define an evaluation framework to assess reviewer recommendations across
endorsement, suppression, preference, and qualification capabilities.

\noindent$\bullet$ We propose \textbf{Pro-MMoE},
a novel framework designed to overcome the limitations of implicit embeddings
and single-objective optimization.
By combining the LLM-Summary and Raw-text branches with a task-adaptive MMoE architecture,
our approach preserves fine-grained expertise, improves interpretability,
and supports distinct confidence and ranking objectives.

\noindent$\bullet$ Experiments demonstrate that Pro-MMoE achieves the best overall performance on the OmniReview benchmark,
while the taxonomy and summaries make reviewer recommendations easier to inspect.

\begin{table*}[t]
\centering
\small
\setlength{\tabcolsep}{3pt}

\renewcommand{\arraystretch}{1.1}

\begin{tabular}{@{}c|c|c|cc|cc|cccc@{}}
\toprule
\multirow{2}{*}{\textbf{Dataset}}
& \multirow{2}{*}{\makecell[c]{\textbf{Review}\\ \textbf{Records}}}
& \multirow{2}{*}{\makecell[c]{\textbf{Review} \\ \textbf{Candidates}}}
& \multicolumn{2}{c|}{\textbf{Profiling}}
& \multicolumn{2}{c|}{\textbf{Labels}}
& \multicolumn{4}{c}{\textbf{Evaluation}}\\
\cmidrule(l){4-11}
&&& 
\textbf{Data Linkage}
& \textbf{ORCID} 
& \textbf{GT.}
& \textbf{Verified}
& \textbf{Endo.}
& \textbf{Supp.}
& \textbf{Pref.}
& \textbf{Qual.} \\
\midrule

NIPS
& 34   
& 190  
& \textcolor{red}{\textbf{\ding{55}}} 
& \textcolor{red}{\textbf{\ding{55}}} 
& \textcolor{red}{\textbf{\ding{55}}} 
& \textcolor{forestgreen}{\textbf{\ding{51}}} 
& \textcolor{forestgreen}{\textbf{\ding{51}}} 
& \textcolor{red}{\textbf{\ding{55}}} 
& \textcolor{red}{\textbf{\ding{55}}} 
& \textcolor{red}{\textbf{\ding{55}}} \\

KDD
& 174  
& 737  
& \textcolor{red}{\textbf{\ding{55}}} 
& \textcolor{red}{\textbf{\ding{55}}} 
& \textcolor{red}{\textbf{\ding{55}}} 
& \textcolor{forestgreen}{\textbf{\ding{51}}} 
& \textcolor{forestgreen}{\textbf{\ding{51}}} 
& \textcolor{red}{\textbf{\ding{55}}} 
& \textcolor{red}{\textbf{\ding{55}}} 
& \textcolor{red}{\textbf{\ding{55}}} \\

exHarmony
& 1204k   
& 1590k  
& \textcolor{forestgreen}{\textbf{\ding{51}}} 
& \textcolor{red}{\textbf{\ding{55}}} 
& \textcolor{red}{\textbf{\ding{55}}} 
& \textcolor{red}{\textbf{\ding{55}}} 
& \textcolor{forestgreen}{\textbf{\ding{51}}} 
& \textcolor{red}{\textbf{\ding{55}}} 
& \textcolor{red}{\textbf{\ding{55}}} 
& \textcolor{red}{\textbf{\ding{55}}} \\

OAG-Bench
& 40k  
& 42k  
& \textcolor{forestgreen}{\textbf{\ding{51}}} 
& \textcolor{red}{\textbf{\ding{55}}} 
& \textcolor{forestgreen}{\textbf{\ding{51}}} 
& \textcolor{red}{\textbf{\ding{55}}} 
& \textcolor{forestgreen}{\textbf{\ding{51}}} 
& \textcolor{red}{\textbf{\ding{55}}} 
& \textcolor{red}{\textbf{\ding{55}}} 
& \textcolor{red}{\textbf{\ding{55}}} \\

ACM-DL Library
& 13k  
& 23k  
& \textcolor{red}{\textbf{\ding{55}}} 
& \textcolor{red}{\textbf{\ding{55}}} 
& \textcolor{red}{\textbf{\ding{55}}} 
& \textcolor{forestgreen}{\textbf{\ding{51}}} 
& \textcolor{forestgreen}{\textbf{\ding{51}}} 
& \textcolor{red}{\textbf{\ding{55}}} 
& \textcolor{red}{\textbf{\ding{55}}} 
& \textcolor{red}{\textbf{\ding{55}}} \\

FRONTIER-RevRec
& 478k 
& 178k 
& \textcolor{red}{\textbf{\ding{55}}} 
& \textcolor{red}{\textbf{\ding{55}}} 
& \textcolor{forestgreen}{\textbf{\ding{51}}} 
& \textcolor{red}{\textbf{\ding{55}}} 
& \textcolor{forestgreen}{\textbf{\ding{51}}} 
& \textcolor{red}{\textbf{\ding{55}}} 
& \textcolor{red}{\textbf{\ding{55}}} 
& \textcolor{red}{\textbf{\ding{55}}} \\

LR-bench (RATE)
& 825
& 406
& \textcolor{red}{\textbf{\ding{55}}}
& \textcolor{red}{\textbf{\ding{55}}}
& \textcolor{red}{\textbf{\ding{55}}}
& \textcolor{forestgreen}{\textbf{\ding{51}}}
& \textcolor{red}{\textbf{\ding{55}}}
& \textcolor{red}{\textbf{\ding{55}}}
& $\dagger$
& \textcolor{red}{\textbf{\ding{55}}} \\

\midrule

OmniReview (Ours) 
& 203k 
& 2288k
& \textcolor{forestgreen}{\textbf{\ding{51}}} 
& \textcolor{forestgreen}{\textbf{\ding{51}}} 
& \textcolor{forestgreen}{\textbf{\ding{51}}} 
& \textcolor{forestgreen}{\textbf{\ding{51}}} 
& \textcolor{forestgreen}{\textbf{\ding{51}}} 
& \textcolor{forestgreen}{\textbf{\ding{51}}} 
& \textcolor{forestgreen}{\textbf{\ding{51}}} 
& \textcolor{forestgreen}{\textbf{\ding{51}}} \\

\bottomrule
\end{tabular}
\caption{Comparison of OmniReview and existing reviewer recommendation datasets.
Data Linkage denotes linking reviewer identities to external scholarly datasets or academic graphs to enrich profiles;
ORCID denotes identity-registry linkage. GT. denotes historical reviewer assignments;
Verified denotes human-assessed or manually checked labels. Endo., Supp., Pref., and Qual. denote endorsement, suppression, preference, and qualification. }
\label{tab:dataset_comparison}
\end{table*}

\section{Related Work}
\subsection{Reviewer Recommendation Datasets}
Table~\ref{tab:dataset_comparison} summarizes the evolution of reviewer recommendation datasets from small benchmarks to large, multidisciplinary resources.
The seminal NIPS dataset~\cite{mimno2007expertise} introduced expert-annotated relevance judgments for $34$ papers and $190$ reviewers. Building on this work, Zhang et al.~\cite{zhang2025chain} combined three existing datasets with a newly annotated KDD dataset containing relevance scores for $3,480$ paper--reviewer pairs. These early datasets remained limited to specific domains. OAG-Bench~\cite{zhang2024oag} expanded disciplinary coverage and linked metadata to a comprehensive scholarly graph. The ACM-DL dataset~\cite{zhang2020multi} used information from the ACM Digital Library to frame paper--reviewer matching as multi-label classification with $1,944$ CCS research field labels. exHarmony~\cite{ebrahimi2025exharmony} used the OpenAlex scholarly graph, treating paper authors as potentially suitable reviewers for related submissions. FRONTIER-RevRec~\cite{peng2025frontier} expanded the scale to $177,941$ reviewers and $478,379$ papers. RATE~\cite{liu-etal-2026-rate} introduces LR-bench, with 1,055 self-reported familiarity ratings across 825 papers and 406 researchers, supporting pairwise expertise-preference evaluation. OmniReview contains $202,756$ review records linked to Open Academic Graph profiles.

\subsection{Recommendation Methods}
CoF~\cite{zhang2025chain} models multiple aspects of paper--reviewer compatibility,
illustrating the value of matching evidence beyond a single undifferentiated similarity score.
RATE~\cite{liu-etal-2026-rate} likewise considers research-topic and methodological compatibility.
In talent search, LLM-derived representations and role-aware MMoE support preference-sensitive ranking~\cite{li2025enhancingtalentsearchranking}.
Building on these representation and multi-task ideas,
Pro-MMoE combines publication-level summaries with raw text for reviewer matching.
Its task-specific heads support candidate confidence and two ranking objectives,
while the discipline taxonomy and readable summaries expose the publication evidence used to characterize candidates.


\section{Dataset Construction}
\subsection{Definition}

We define OmniReview as $\mathcal{D} = (\mathcal{P}, \mathcal{A}, \mathcal{R})$, where $\mathcal{P}$ is the publication collection, $\mathcal{A}$ is the scholar set, and $\mathcal{R}$ is the set of peer-review records.
Construction comprises three phases: cross-source matching using publication titles and co-author evidence, discipline taxonomy construction, and task formulation with potential and unqualified candidates.

\subsection{Cross-source Matching}

The three data sources provide complementary information.
Cross-source matching follows four steps to support reliable scholar identification: (1) data cleaning; (2) publication matching; (3) scholar matching; and (4) verification.

\begin{figure*}[t]
    \centering
    \includegraphics[width=1\linewidth]{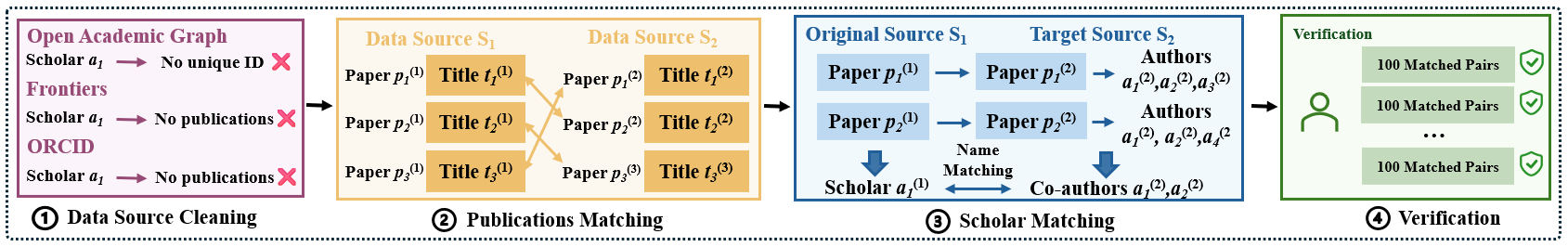}
    \caption{Cross-source disambiguation pipeline linking Frontiers, OAG, and ORCID records through data cleaning, publication matching, scholar matching, and verification.}
    \label{fig:disambiguation}
\end{figure*}

\textbf{(1) Data Cleaning.} We check scholar and publication uniqueness, remove OAG and ORCID author records without unique persistent identifiers, and exclude Frontiers reviewers without publication records. These records lack the identifiers or scholarly evidence needed for reliable disambiguation and expertise assessment.

\textbf{(2) Publication Matching.} We match articles across sources by normalized titles. For a publication $p$, $\mathcal{N}(p)=(w_1,\ldots,w_m)$ lowercases the title, removes punctuation, and splits it into words. Publications $p^{(1)}\in\mathcal{P}^{(1)}$ and $p^{(2)}\in\mathcal{P}^{(2)}$ match only when their normalized word sequences are identical, eliminating differences in capitalization and punctuation. Algorithm~\ref{alg:title_word_matching} provides the procedure.

\textbf{(3) Scholar Matching.} We normalize name variants, including Roman numerals and pinyin, into standardized Latin-script forms. For scholar $a$ in source $S_i$, we locate publications in $S_j$ matching those in $P_i^{(a)}$, then obtain candidate identities $A'$ from the intersection of the matched publications' author sets. A two-stage procedure selects the candidate with the greatest normalized name-token overlap; ties or zero overlap are resolved by exact agreement of name-token initials. Algorithm~\ref{alg:scholar_matching} details this procedure.

\textbf{(4) Verification.} 
To further evaluate the accuracy of scholar matching, we conducted five rounds of manual auditing on randomly sampled matched pairs, with 100 pairs inspected in each round. Across all audited samples, the matched records were consistently correct in key fields, including author information and abstracts, supporting the reliability of the cross-source alignment process.


\subsection{Construction of the Discipline Taxonomy}

Our dataset contains 109,579,745 publications and 34,855,513 researcher profiles, and manual annotation of article-to-discipline assignments is infeasible.
To facilitate profile retrieval, we organize the corpus with a hierarchical discipline taxonomy $\mathcal{T}=\{L_1,L_2,L_3\}$, where $L_1$ denotes broad research domains and $L_3$ corresponds to fine-grained research topics.
In our implementation, the taxonomy structure itself is based on an existing disciplinary standard shown in Appendix \ref{app:discipline_taxonomy}.
We use Qwen3-Embedding-4B~\cite{qwen3embedding} to encode each article's title and abstract and assign it to the subject node with the highest semantic similarity. Each scholar $a \in \mathcal{A}$ then receives a subject profile formed by the union of their publication assignments,
$
\hat{l}_i^{(a)} = \bigcup_{p \in P_a} \hat{l}_i^{(p)},
$
where $P_a$ denotes the publication set of scholar $a$, and $\hat{l}_i^{(p)}$ is the subject node assigned to publication $p$ at taxonomy level $i$. Assignment accuracy is reported in Appendix~\ref{app:discipline_verficiation}.

\subsection{Task Statement}
\label{sec:task}

\textbf{$\bullet$ Candidate Definition.}
Reviewer candidates should have relevant disciplinary backgrounds.
We construct the candidate pool using discipline-taxonomy retrieval and automatic screening rules
informed by the reviewer-selection experience of several editors of IEEE and ACM publications
(Appendix~\ref{app:candidate_screening}).
Within this pool, publication evidence in the specific target discipline defines domain qualification.
For paper $p$, let $c_p$ denote its specific target category and $c^*$ a related neighboring category.
\textbf{Ground-truth reviewers} $\mathcal{G}_p$ are scholars who actually reviewed $p$ in the historical record.
\textbf{Potential reviewers} $\mathcal{P}_p$ are nonhistorical candidates with substantial publication records in $c^*$
and at least one publication in $c_p$.
\textbf{Unqualified candidates} $\mathcal{U}_p$ are nonhistorical candidates with substantial publication records in $c^*$ but no publication in $c_p$.
This publication-based construction grounds candidate membership in
identifiable research records and applies a consistent criterion at scale.
It reduces the need for specialists to annotate individual paper--reviewer pairs
and limits the influence of third-party annotators' subjective expertise judgments.

\textbf{$\bullet$ Evaluation Tasks.}
We evaluate four complementary capabilities:
\textbf{(1) Endorsement}:
We evaluate endorsement through the predicted suitability scores assigned to historical reviewers in $\mathcal{G}_p$.
\textbf{(2) Suppression}:
This task evaluates a recommendation system's ability to suppress scholars who lack domain-specific expertise despite apparent disciplinary proximity.
\textbf{(3) Preference}:
For each paper $p$, we compare historical reviewers $\mathcal{G}_p$ with potential reviewers $\mathcal{P}_p$. Both sets possess topic-relevant expertise, but only the former were historically assigned.
This task aims to evaluate whether a system can mimic real editorial preference by ranking historically selected reviewers above other qualified candidates.
\textbf{(4) Qualification}:
To address false negatives from historical assignments, we require potential reviewers $\mathcal{P}_p$ to rank above unqualified candidates $\mathcal{U}_p$, testing whether a system recognizes qualified but previously unassigned experts.

\begin{figure*}[t]
    \centering
    \includegraphics[width=0.9\linewidth]{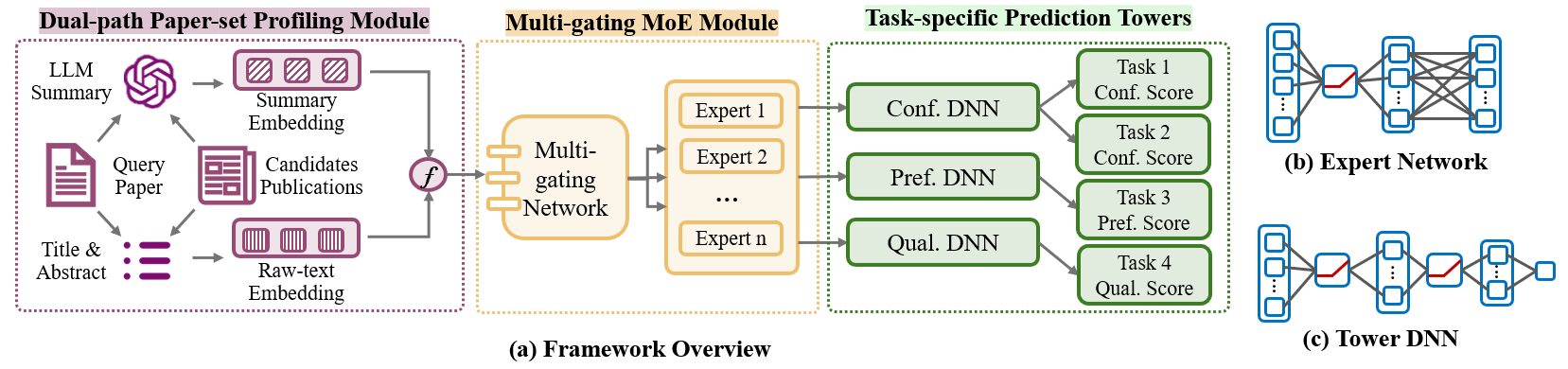}
    \caption{Pro-MMoE: LLM-Summary and Raw-text branches encode the target paper and candidate publications. Fused interaction features pass through shared experts and task-specific gates to confidence (Conf.), preference (Pref.), and qualification (Qual.) towers.}
    \label{fig:mmoeframework}
\end{figure*}

\section{Method}

We propose a dual-path multi-gate mixture-of-experts framework for reviewer recommendation.
The following subsections describe the framework's three components: dual-path profiling, an MMoE module, and prediction towers (Figure~\ref{fig:mmoeframework}).

\subsection{Dual-path Paper-set Profiling Module}

We represent each reviewer using two complementary views derived from publication-level summaries and original titles and abstracts. The Summary Branch captures high-level research content, while the Raw-text Branch retains fine-grained textual details. Both branches use cached publication embeddings.

For each branch $b \in \{s,r\}$, a mean-pooling reviewer encoder independently aggregates the candidate's publication embeddings into a reviewer representation $\mathbf{r}^{(b)}$. The target paper representation $\mathbf{p}^{(b)}$ lies in the same latent space.
We then construct pairwise interaction features by concatenating the normalized paper representation, the normalized reviewer representation, their absolute difference, their element-wise product, and cosine similarity:

\begin{equation}
\begin{aligned}
\phi^{(b)} = \big[\,& \tilde{\mathbf{p}}^{(b)},\ \tilde{\mathbf{r}}^{(b)},\
\lvert \tilde{\mathbf{p}}^{(b)} - \tilde{\mathbf{r}}^{(b)} \rvert,\\
& \tilde{\mathbf{p}}^{(b)} \odot \tilde{\mathbf{r}}^{(b)},\
\cos\big(\tilde{\mathbf{p}}^{(b)},\tilde{\mathbf{r}}^{(b)}\big)\,\big],\\
\mathbf{h}_{\text{fuse}} &= f_{\text{fusion}}\left([\phi^{(s)},\phi^{(r)}]\right).
\end{aligned}
\end{equation}

where $\tilde{\mathbf{p}}^{(b)}$ and $\tilde{\mathbf{r}}^{(b)}$ denote layer-normalized representations. The interaction features from the Summary and Raw-text branches are concatenated and passed through a fusion MLP to obtain the final shared representation.

\subsection{Multi-gate Mixture-of-Experts Module}

The fused representation $\mathbf{h}_{\text{fuse}}$ is fed into an MMoE module~\cite{ma2018modeling} with $n$ shared feed-forward expert networks $\{E_i\}_{i=1}^{n}$. Here, $k\in\{0,1,2\}$ indexes the confidence, preference, and qualification objectives, respectively. For each objective, a gate $G^{(k)}$ produces a routing distribution over the shared experts, yielding a weighted combination of expert outputs:
\begin{equation}
\begin{aligned}
\mathbf{g}^{(k)} &= \mathrm{softmax}\!\left(G^{(k)}(\mathbf{h}_{\text{fuse}})\right),\\
\hat{\mathbf{h}}^{(k)} &= \sum_{i=1}^{n}g_i^{(k)}E_i(\mathbf{h}_{\text{fuse}}).
\end{aligned}
\end{equation}


\subsection{Task-specific Prediction Towers and Training Objectives}

Three prediction towers operate on the corresponding representations $\hat{\mathbf{h}}^{(k)}$ and support four evaluation tasks. The confidence tower is shared by Tasks 1 and 2, which evaluate endorsement and suppression through suitability scores. The preference tower ranks historical reviewers above other qualified candidates (Task 3), while the qualification tower ranks potential reviewers above unqualified candidates (Task 4).

For Tasks 1 and 2, we optimize binary cross-entropy on logits with hard-negative reweighting, so that unqualified candidates assigned high confidence receive larger penalties:
\begin{equation}
\begin{aligned}
\mathcal{L}_{conf} &= \mathrm{BCEWithLogits}(s,y;w),\\
w &=
\begin{cases}
1+\alpha \sigma(s)^{\gamma}, & y=0,\\
1, & y=1.
\end{cases}
\end{aligned}
\end{equation}

For Tasks 3 and 4, we use a query-level pairwise logistic ranking loss. Given a query group, let $\mathcal{P}=\{(i,j)\mid y_i>y_j\}$ denote the set of valid preference pairs. The ranking objective is
\begin{equation}
\mathcal{L}_{\mathrm{rank}}
=
\frac{1}{|\mathcal{P}|}
\sum_{(i,j)\in\mathcal{P}}
\log\!\left(1+\exp(s_j-s_i)\right).
\end{equation}

The training objectives combine the tower losses with gate regularization, with tower weights activated according to the training stage:
\begin{equation}
\begin{aligned}
\mathcal{L} ={}& \lambda_0\mathcal{L}_{conf}+\lambda_1\mathcal{L}_{pref}
+\lambda_2\mathcal{L}_{qual}\\
&-\lambda_{\mathrm{ent}}\mathcal{H}(\mathbf{g})
+\lambda_{\mathrm{lb}}\mathrm{Std}(\bar{\mathbf{g}}),
\end{aligned}
\label{eq:training_objective}
\end{equation}
where $\mathcal{L}_{pref}$ and $\mathcal{L}_{qual}$ are instantiated by $\mathcal{L}_{\mathrm{rank}}$, $\mathcal{H}(\mathbf{g})=-\sum_i g_i\log g_i$ encourages non-collapsed routing, and $\mathrm{Std}(\bar{\mathbf{g}})$ penalizes imbalanced expert utilization.

Training proceeds in two stages. Stage 1 activates only the confidence loss, with $(\lambda_0,\lambda_1,\lambda_2)=(1,0,0)$, and trains the reranker modules used by that objective. Stage 2 starts from the best Stage 1 checkpoint and activates the preference and qualification losses, with weights $(0,1,1)$. In the reported configuration, the shared modules and confidence tower are frozen; only the ranking gates, towers, and scorers are updated. The summarization and embedding models provide cached inputs and are not optimized during reranker training.

\section{Experiments}

\subsection{Experimental Settings}
\textbf{$\bullet$ Benchmark Dataset.}
Each review record contains paper, reviewer, and editor metadata, together with a category from our discipline taxonomy. An example record and usage instructions are provided in Appendices~\ref{app:omni_example} and~\ref{app:dataset_usage}.
CoF~\citep{zhang2025chain} evaluates 388 papers, while exHarmony~\citep{ebrahimi2025exharmony} tests 10K papers with 30K candidates.
Our benchmark comprises 812 review records, 14K candidate reviewers, and 120K papers, selected through stratified sampling over discipline categories.
Scholars with prior peer-review experience manually verified potential and unqualified candidates' domain expertise using their publication records and the target papers' titles and abstracts.
The remaining records are split into training and validation sets.
All supervised model training and score calibration use the same training split.
Due to LLM summarization costs, Pro-MMoE uses a discipline-stratified subset of $8,427$ training records in the main experiments; the data-volume ablation varies its size.

\textbf{$\bullet$ Evaluation Metrics.}
We evaluate the proposed framework on the four tasks introduced in Section~\ref{sec:task}. 
For Task 1, we report real reviewer confidence (RRC), defined as the mean predicted suitability score assigned to historical reviewers.
For Task 2, we report unqualified candidate confidence (UCC), defined as the mean predicted suitability score assigned to unqualified candidates; lower values indicate stronger suppression.
For Tasks 3 and 4, we evaluate ranking quality using mean average precision (MAP) and normalized discounted cumulative gain (NDCG).
RRC and UCC jointly assess whether a method endorses historical reviewers while suppressing unqualified candidates; high RRC alone does not establish effective discrimination.

\begin{table*}[t]
\centering
\small
\setlength{\tabcolsep}{4pt}

\begin{tabular}{
l l
S[table-format=1.4]
S[table-format=1.4]
S[table-format=1.4]
S[table-format=1.4]
S[table-format=1.4]
S[table-format=1.4]
}
\toprule
\multirow{2}{*}{\textbf{Method}} &
\multirow{2}{*}{\textbf{Category}} &
\multicolumn{1}{c}{\textbf{Task 1}} &
\multicolumn{1}{c}{\textbf{Task 2}} &
\multicolumn{2}{c}{\textbf{Task 3}} &
\multicolumn{2}{c}{\textbf{Task 4}} \\
\cmidrule(lr){3-3} \cmidrule(lr){4-4} \cmidrule(lr){5-6} \cmidrule(lr){7-8}
&
&
\multicolumn{1}{c}{\textbf{RRC}$\uparrow$} &
\multicolumn{1}{c}{\textbf{UCC}$\downarrow$} &
\multicolumn{1}{c}{\textbf{MAP}$\uparrow$} &
\multicolumn{1}{c}{\textbf{NDCG}$\uparrow$} &
\multicolumn{1}{c}{\textbf{MAP}$\uparrow$} &
\multicolumn{1}{c}{\textbf{NDCG}$\uparrow$} \\
\midrule

TF-IDF & Lexical
& 0.8540 & 0.3828 & 0.6712 & 0.7679
& 0.5282 & 0.7126 \\

Dual-tower & Bi-encoder
& 0.8177 & 0.4428 & 0.4217 & 0.5580
& 0.7210 & 0.8555 \\

RGCN & Graph
& 0.6813 & 0.3841 & 0.5170 & 0.6501
& 0.6835 & 0.8362 \\

SciBERT & Embedding
& 0.7287 & 0.4675 & 0.5590 & 0.6853
& 0.5858 & 0.7634 \\

SPECTER2 & Embedding
& 0.8345 & 0.4808 & 0.6455 & 0.7479
& 0.6378 & 0.8011 \\

BGE-M3 & Embedding
& 0.8628 & 0.5276 & 0.6071 & 0.7204
& 0.6363 & 0.7982 \\

Qwen3-4B-Emb. & Embedding
& 0.8539 & 0.4516 & 0.6064 & 0.7187
& 0.6596 & 0.8141 \\

CoF & Multi-Emb.
& \uline{0.8768} & \uline{0.3681} & \uline{0.6826} & \uline{0.7741}
& \textbf{0.7669} & \uline{0.8788} \\

LLM-as-Judge & LLM-based
& 0.6920 & 0.6216 & 0.3933 & 0.5592
& 0.4716 & 0.6735 \\

Pro-MMoE (ours) & LLM+MMoE
& \textbf{0.8851} & \textbf{0.1897} & \textbf{0.7062} & \textbf{0.7943}
& \uline{0.7408} & \textbf{0.8805} \\

\bottomrule
\end{tabular}
\caption{Performance comparison across the four tasks on OmniReview. Emb. denotes embedding.}
\label{tab:reviewer_recommendation_performance}
\end{table*}

\textbf{$\bullet$ Methods.}
We compare traditional retrieval, dense embedding, graph-based, and LLM-based approaches. The baselines include: (1) a \textbf{TF-IDF-based similarity method}~\citep{ramos2003using} for lexical matching; (2) a \textbf{dual-tower model}~\citep{huang2013learning} with separate paper and reviewer encoders; (3) a \textbf{Relational Graph Convolutional Network (RGCN)}~\citep{schlichtkrull2018modeling} using heterogeneous scholarly graph structure; (4) \textbf{embedding-based models}, including SciBERT~\citep{beltagy2019scibertpretrainedlanguagemodel}, SPECTER2~\citep{Singh2022SciRepEvalAM}, BGE-M3~\citep{bge-m3}, Qwen3-Embedding-4B~\citep{qwen3embedding}, and Chain-of-Factors (CoF)~\citep{zhang2025chain}, which capture semantic, topic, or citation information; and (5) \textbf{an LLM-as-Judge approach}~\citep{karan-etal-2025-dataset} for direct suitability assessment. Baseline implementations, training details, hyperparameters, and Pro-MMoE's computational costs are described in Appendix~\ref{app:baseline}.

\subsection{Results and Analysis}
\label{sec:results}


Table~\ref{tab:reviewer_recommendation_performance} shows that \textbf{Pro-MMoE} achieves the best scores on five of the six reported metrics. Its largest improvement over \textbf{CoF}, the strongest competing baseline, is in suppressing unqualified candidates: UCC decreases from 0.3681 to 0.1897, a relative reduction of approximately 48.5\%. Meanwhile, RRC increases from 0.8768 to 0.8851, showing that stronger suppression is accompanied by greater endorsement of historical reviewers.

For historical reviewer preference (Task~3), Pro-MMoE improves MAP from CoF's 0.6826 to 0.7062 and NDCG from 0.7741 to 0.7943, indicating better reproduction of historical assignments. For qualification ranking (Task~4), it obtains the highest NDCG of 0.8805, while CoF achieves higher MAP (0.7669 versus 0.7408). Thus, neither method dominates both qualification metrics.

\textbf{TF-IDF} outperforms the individual dense embedding models on Task~3, showing the value of lexical overlap for historical preference. These embedding models all outperform TF-IDF on Task~4 MAP and NDCG, supporting semantic matching for identifying potential reviewers. \textbf{RGCN} scores below CoF and Pro-MMoE on all six metrics.

The tested \textbf{LLM-as-Judge} configuration has an RRC of 0.6920 and a UCC of 0.6216, with weak ranking scores. Direct suitability assessment thus provides limited discrimination in this setting.

\subsection{Ablation Study}

\begin{figure*}[t]
    \centering

    \begin{minipage}[t]{0.48\linewidth}
        \centering
        \includegraphics[width=\linewidth]{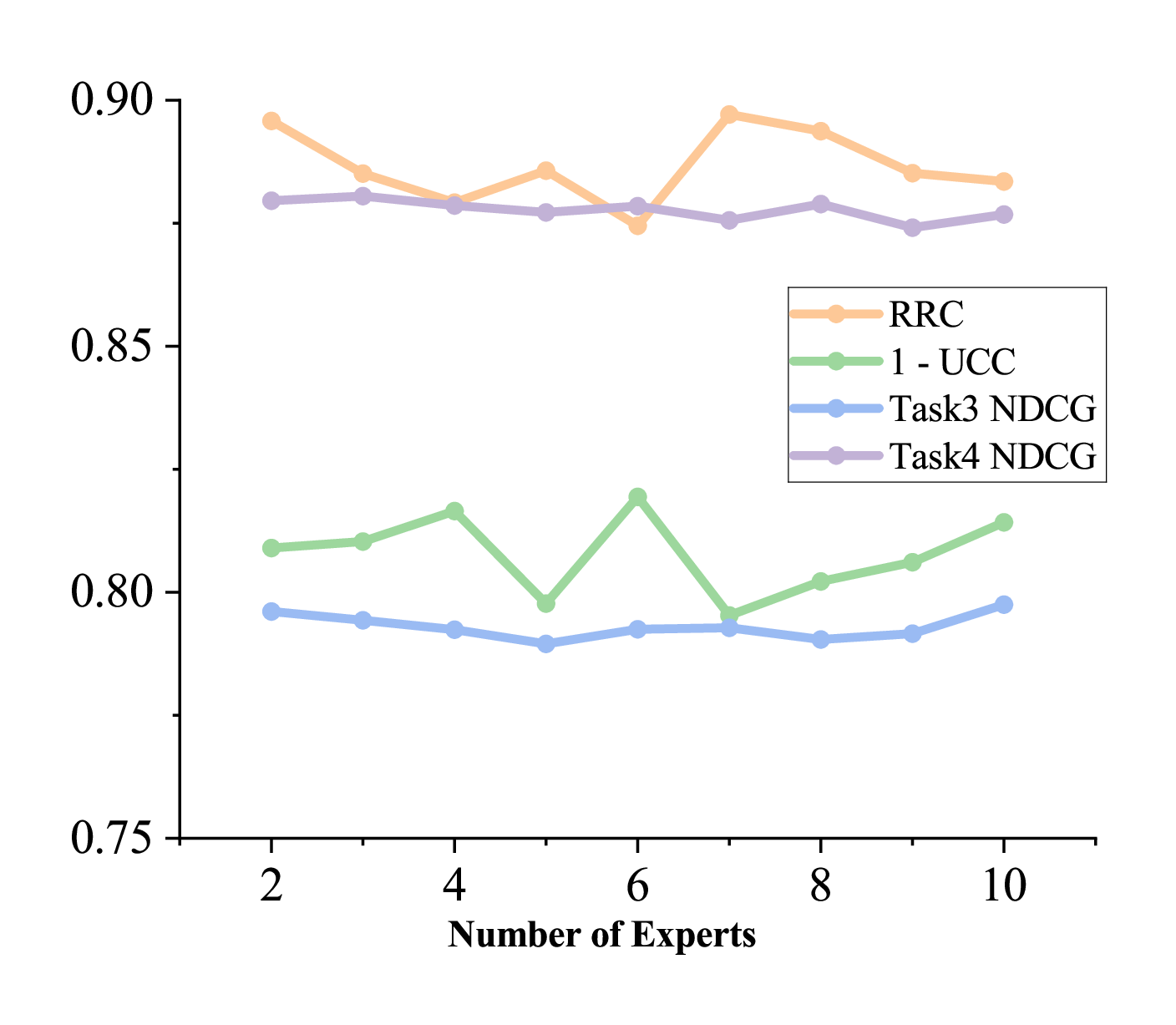}
        \centerline{\small (a) Number of experts}
    \end{minipage}
    \hfill
    \begin{minipage}[t]{0.48\linewidth}
        \centering
        \includegraphics[width=\linewidth]{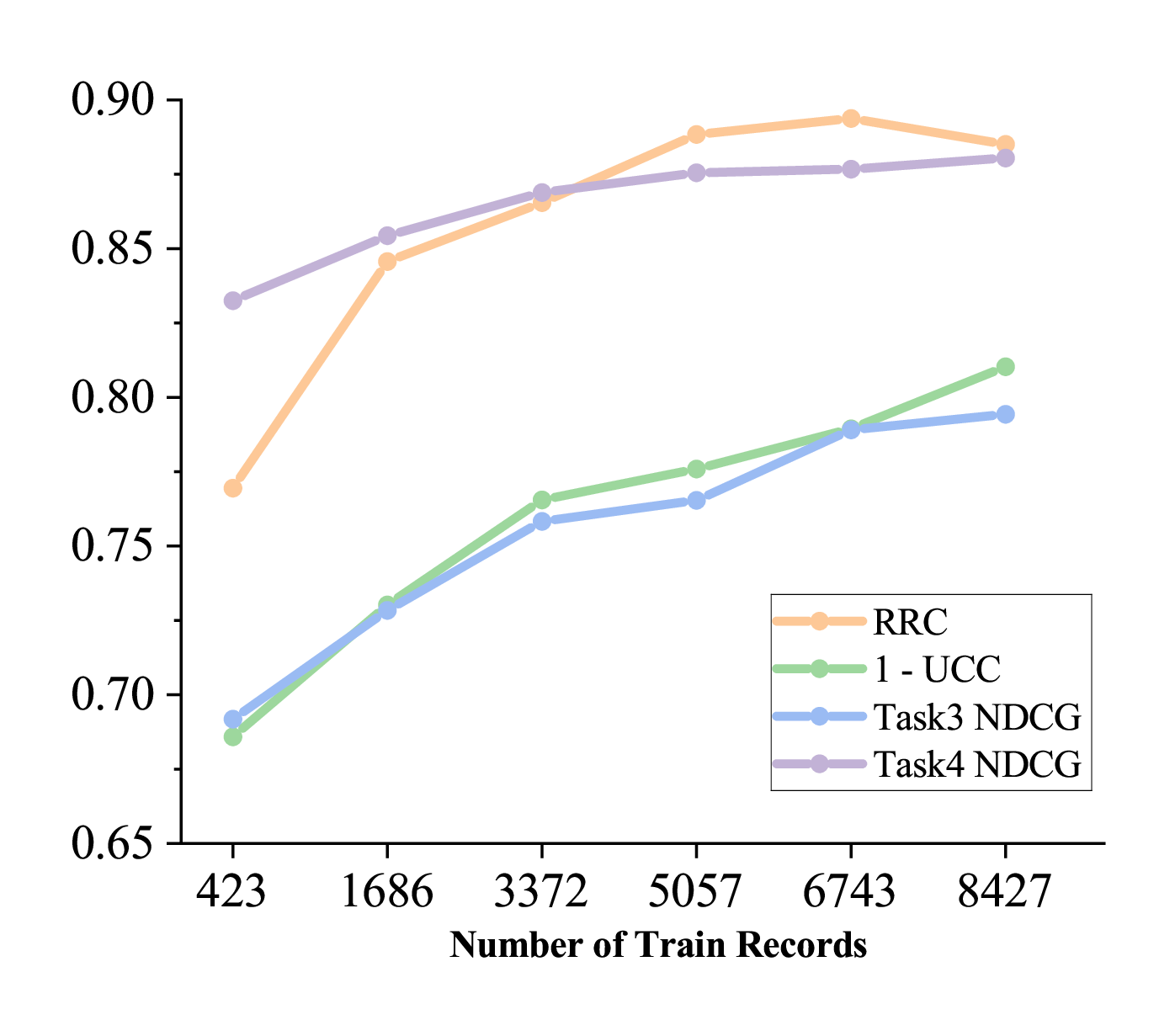}
        \centerline{\small (b) Training data volume}
    \end{minipage}

    \caption{Effect of the number of MMoE experts (a) and training data volume (b). The default uses three experts; performance remains competitive with $3,372$ training records.}
    \label{fig:expert_data_ablation}
\end{figure*}

\begin{table}[t]
\centering

\resizebox{\columnwidth}{!}{%
\begin{tabular}{lcccc}
\toprule
\multirow{2}{*}{\textbf{Method}} 
& \multicolumn{1}{c}{\textbf{Task 1}} 
& \multicolumn{1}{c}{\textbf{Task 2}} 
& \multicolumn{1}{c}{\textbf{Task 3}} 
& \multicolumn{1}{c}{\textbf{Task 4}} \\
& \textbf{RRC} $\uparrow$ 
& \textbf{UCC} $\downarrow$ 
& \textbf{NDCG} $\uparrow$ 
& \textbf{NDCG} $\uparrow$ \\
\midrule
Pro-MMoE & \uline{0.8851} & \textbf{0.1897} & \textbf{0.7943} & \textbf{0.8805} \\
\hline
w/o Sum. Branch & 0.8653 & \uline{0.2023} & 0.7773 & 0.8801 \\
w/o Raw. Branch & 0.8719 & 0.2278 & 0.7681 & 0.8786 \\
\midrule
w/ Reviewer Sum. & \textbf{0.9292} & 0.8592 & 0.5461 & 0.8751 \\
w/o MMoE & 0.7487 & 0.2771 & \uline{0.7902} & \uline{0.8802} \\
\bottomrule
\end{tabular}
}
\caption{Component ablations and reviewer-level summarization. Bold and underlined values denote the best and second-best scores, respectively.}
\label{tab:llm_ablation}
\end{table}

\textbf{$\bullet$ Framework Components.}
Table~\ref{tab:llm_ablation} shows that removing either input branch reduces performance on all four reported metrics. Without summaries, Task~3 NDCG falls from 0.7943 to 0.7773. Removing raw text causes a larger decrease to 0.7681 and raises UCC from 0.1897 to 0.2278. These changes support combining semantic abstraction with textual details to distinguish closely related candidates. Task~4 NDCG is less sensitive to either removal.

The reviewer-level summarization variant (w/ Reviewer Sum.) replaces mean-pooled publication-level summaries with one profile summary. RRC increases to 0.9292, but UCC rises sharply to 0.8592 and Task~3 NDCG falls to 0.5461. High endorsement scores therefore do not imply effective discrimination. A possible explanation is that a single summary emphasizes broad interests while obscuring publication-specific expertise differences.

Removing MMoE reduces RRC to 0.7487 and raises UCC to 0.2771, while Task~3 and Task~4 NDCG remain close to the full model, at 0.7902 and 0.8802. Its benefit in this ablation is therefore concentrated in confidence estimation, showing that component contributions differ across objectives.

\begin{table}[t]
\centering
\small
\setlength{\tabcolsep}{2pt}
\resizebox{\columnwidth}{!}{%
\begin{tabular}{lcccc}
\toprule
\textbf{Backbone (Param.)}
& \makecell{\textbf{Task 1}\\\textbf{RRC} $\uparrow$}
& \makecell{\textbf{Task 2}\\\textbf{UCC} $\downarrow$}
& \makecell{\textbf{Task 3}\\\textbf{NDCG} $\uparrow$}
& \makecell{\textbf{Task 4}\\\textbf{NDCG} $\uparrow$} \\
\midrule
\multicolumn{5}{l}{\textit{Summarization backbone}} \\
Qwen3-30B-A3B (30B) & \textbf{0.8851} & \uline{0.1897} & \textbf{0.7943} & \textbf{0.8805} \\
Gemma4-4B (4B) & 0.8505 & \textbf{0.1637} & 0.7859 & 0.8787 \\
Qwen3.5-4B (4B) & \uline{0.8811} & 0.1977 & \uline{0.7911} & \uline{0.8803} \\
\midrule
\multicolumn{5}{l}{\textit{Embedding backbone}} \\
Qwen3-Emb. (4B) & \textbf{0.8851} & \textbf{0.1897} & \textbf{0.7943} & \textbf{0.8805} \\
BGE-M3 (0.6B) & 0.8504 & 0.2160 & 0.7582 & 0.8709 \\
\bottomrule
\end{tabular}%
}

\caption{Comparison of summarization and embedding backbones. Emb. denotes Embedding-4B.}
\label{tab:backbone_embedding_compare}
\end{table}

\textbf{$\bullet$ Backbones.}
Table~\ref{tab:backbone_embedding_compare} shows that smaller summarization backbones remain competitive. Qwen3.5-4B obtains Task~3 and Task~4 NDCG scores of 0.7911 and 0.8803, close to the default's. Gemma4-4B has the lowest UCC of 0.1637 but lower RRC and ranking scores, so greater model size is not uniformly beneficial.

Replacing Qwen3-Embedding-4B with BGE-M3 reduces Task~3 NDCG to 0.7582 and Task~4 NDCG to 0.8709, with lower RRC and higher UCC. Historical preference ranking is more sensitive to this substitution than qualification ranking.

\textbf{$\bullet$ Number of Experts.}
Other tested hyperparameters had relatively small effects, so we focus on the number of MMoE experts.
Figure~\ref{fig:expert_data_ablation}(a) shows that RRC remains well above UCC across configurations; the plot uses $1-\mathrm{UCC}$ so that higher values consistently indicate better performance. Ranking metrics vary within narrow ranges, while confidence metrics fluctuate more. More experts do not consistently improve all objectives, so we retain three as a compact default.

\textbf{$\bullet$ Data Volume.}
We vary the discipline-stratified training subset from $423$ to $8,427$ records. Figure~\ref{fig:expert_data_ablation}(b) shows that more training data generally improves performance, although RRC is not strictly monotonic. Task~4 NDCG remains comparatively strong at low data volumes, whereas historical preference and suppression improve more noticeably with additional records. With $3,372$ records, the model remains competitive but does not match the full-data configuration on every objective. These trends indicate useful matching behavior under limited supervision and continued benefits from additional training data.

\FloatBarrier
\section{Conclusion}
\label{sec:conclusion}

OmniReview links Frontiers review records with OAG and ORCID profiles, yielding $202,756$ verified records and $2,287,697$ candidates, including $150,287$ historical reviewers. Four tasks evaluate endorsement, suppression, preference, and qualification. Pro-MMoE combines LLM-Summary and Raw-text branches with task-adaptive gating and two-stage training for confidence estimation and ranking, achieving the best overall performance. These resources support transparent, efficient, and equitable peer review.

\section*{Limitations}

OmniReview is constructed from Frontiers peer-review records linked to OAG and ORCID. Its coverage therefore depends on these sources, and results may not generalize to other publishers or reviewing workflows. Researchers without sufficiently complete publication records may be underrepresented. Cross-source identity resolution and embedding-based discipline assignment can introduce errors; the manual verification of taxonomy assignment reported in Appendix~\ref{app:discipline_verficiation} does not establish that every candidate label is correct.

OmniReview supports reviewer recommendation using integrated cross-source scholarly data, with candidate profiles and publication-based qualification labels constructed from the same database snapshot. Historical reviewer labels record observed paper--reviewer relationships and remain unchanged by subsequent publication updates. Our evaluation assesses matching within this shared information setting and does not aim to reconstruct reviewer profiles at individual historical review dates.

Historical assignments and publication-based expertise are proxies for reviewer suitability. They do not establish reviewer availability, willingness to review, or the absence of conflicts of interest. Ranking historical reviewers above other qualified candidates models past editorial choices, which may themselves contain biases.

The Summary Branch depends on LLM-generated summaries of titles and abstracts, which may omit expertise details or introduce unsupported interpretations. Main comparisons report single-run results without variance estimates or significance tests, so small differences should be interpreted cautiously. Pro-MMoE also does not achieve the best score on every metric, as discussed in Section~\ref{sec:results}. Finally, generating summaries and embeddings incurs substantial preprocessing cost even when subsequent scoring uses cached representations.

\section*{Ethical Considerations}

Open Academic Graph (OAG) is released under the Open Data Commons Attribution (ODC-BY) license.
ORCID provides its Public Data File under the CC0 1.0 Public Domain Dedication.
Article metadata from the Frontiers open-access platform is distributed under Creative Commons CC0 terms.
All data are used in accordance with their respective licenses.

OmniReview and Pro-MMoE are intended to support editors by providing more transparent and evidence-based reviewer recommendations. However, a potential negative societal impact is over-reliance on automated recommendation scores. If editors treat model outputs as final decisions rather than auxiliary evidence, human judgment may be weakened, and unfair reviewer assignments may occur.

\bibliography{reference}

\clearpage
\appendix
\section{Details of Data Source Alignment}
\label{app:data_source_alignment}

Algorithm~\ref{alg:title_word_matching} performs cross-source publication alignment by conservatively matching normalized titles through exact token-level comparison, thereby producing high-precision publication correspondences as anchors for subsequent entity linking. Algorithm~\ref{alg:scholar_matching} then resolves scholar identities across sources by restricting candidates to authors of the matched publications and disambiguating names through normalized token overlap and initials consistency checks.

\begin{algorithm*}[t]
\centering

\caption{Publication Matching for Cross-Source Publications}
\label{alg:title_word_matching}
\begin{algorithmic}[1]
\Require Two publication titles $t^{(1)}$ and $t^{(2)}$ from publications
\Statex \hspace{\algorithmicindent} $P^{(1)}$ and $P^{(2)}$ in different data sources
\Statex \hspace{\algorithmicindent} $S^{(1)}$ and $S^{(2)}$
\Ensure Boolean match result indicating whether the titles represent the same publication

\Function{NormalizeTitle}{$title$}
    \State $clean\_title \gets \textsc{remove\_punctuation}(title)$
    \State $normalized \gets \textsc{to\_lowercase}(clean\_title)$
    \State \Return $\textsc{split\_into\_words}(normalized)$
\EndFunction

\State $words^{(1)} \gets \Call{NormalizeTitle}{t^{(1)}}$
\State $words^{(2)} \gets \Call{NormalizeTitle}{t^{(2)}}$

\If{$\textsc{length}(words^{(1)}) \neq \textsc{length}(words^{(2)})$}
    \State \Return false
\EndIf

\For{$k = 1$ to $\textsc{length}(words^{(1)})$}
    \If{$words^{(1)}[k] \neq words^{(2)}[k]$}
        \State \Return false
    \EndIf
\EndFor

\State \Return true
\end{algorithmic}
\end{algorithm*}

\begin{figure}[t]
\begin{enumerate}
    \item \textbf{Mathematics}
    \item \textbf{Information Science and System Science}
    \begin{enumerate}
        \item Basic Disciplines of Information Science and System Science
        \begin{itemize}
            \item Information Theory
            \item Cybernetics
            \item System Theory
        \end{itemize}

        \item Systemology
        \begin{itemize}
            \item Chaos
            \item General System Theory
            \item Dissipative Structure Theory
            \item Synergetics
        \end{itemize}

        \item Control Theory
        \begin{itemize}
            \item Large System Theory
            \item System Identification
            \item State Estimation
            \item Robust Control
        \end{itemize}
    \end{enumerate}
    \item \textbf{Mechanics}
    \item \textbf{Physics}
    \item \textbf{Chemistry}
    \item \textbf{Computer Science and Technology}
    \item \textbf{Statistics}
\end{enumerate}
\caption{Example of the discipline taxonomy used in our benchmark.}
\label{fig:taxonomy_example}
\end{figure}

\section{Details of the Discipline Taxonomy}
\label{app:discipline_taxonomy}
\textbf{Overview.} Our discipline taxonomy is constructed based on the Classification and code of disciplines\footnote{\url{https://openstd.samr.gov.cn/bzgk/gb/newGbInfo?hcno=4C13F521FD6ECB6E5EC026FCD779986E&refer=outter}}.
Due to space constraints, only a subset of the complete discipline tree is shown in Figure~\ref{fig:taxonomy_example}. The full classification tree containing all disciplines is available in the code repository.

\textbf{Verification.}
\label{app:discipline_verficiation}
To assess the reliability of automatic taxonomy assignment, we conducted five rounds of scholar verification. As shown in Table~\ref{tab:taxonomy_validation}, the mean relevance rate is 78.44\%, with a sample standard deviation of 1.99 percentage points and a coefficient of variation of 2.54\%. These results indicate that the taxonomy population procedure is reasonably accurate and stable at large scale.


\begin{algorithm*}[t]
\centering

\caption{Cross-Source Scholar Matching with Optimal Token Matching}
\label{alg:scholar_matching}
\begin{algorithmic}[1]
\Require Scholar $a$ from source $S_i$ with publication set $P_i^{(a)}$
\Statex \hspace{\algorithmicindent} Publication set $P_j$ from source $S_j$ containing
\Statex \hspace{\algorithmicindent} publications with titles matching those in $P_i^{(a)}$
\Ensure Best matched scholar $a'$ from $S_j$, or $\varnothing$ if no match is found

\Function{Normalize}{$name$}
    \State $clean \gets \textsc{ConvertRomanNumerals}(name)$
    \State $clean \gets \textsc{TransliteratePinyin}(clean)$
    \State \Return $clean$
\EndFunction

\State $\mathcal{A}' \gets \bigcap_{p \in P_j} \textsc{GetAuthors}(p)$

\State $name_a \gets \Call{Normalize}{a.\text{name}}$
\State $words^{(a)} \gets \textsc{split\_into\_words}(name_a)$
\State $best\_candidate \gets \varnothing$
\State $max\_score \gets 0$

\For{each $a' \in \mathcal{A}'$}
    \State $name_{a'} \gets \Call{Normalize}{a'.\text{name}}$
    \State $words^{(a')} \gets \textsc{split\_into\_words}(name_{a'})$
    \If{$\lvert words^{(a)} \cap words^{(a')} \rvert > max\_score$}
        \State $max\_score \gets \lvert words^{(a)} \cap words^{(a')} \rvert$
        \State $best\_candidate \gets \{a'\}$
    \ElsIf{$\lvert words^{(a)} \cap words^{(a')} \rvert = max\_score$}
        \State $best\_candidate \gets best\_candidate \cup \{a'\}$
    \EndIf
\EndFor

\If{$max\_score > 0 \land \lvert best\_candidate \rvert = 1$}
    \State \Return the only element in $best\_candidate$
\EndIf

\For{each $a' \in best\_candidate$}
    \State $name_{a'} \gets \Call{Normalize}{a'.\text{name}}$
    \State $initials_a \gets \{w[0] \mid w \in words^{(a)}\}$
    \State $initials_{a'} \gets \{w[0] \mid w \in words^{(a')}\}$
    \If{$initials_a = initials_{a'}$}
        \State \Return $a'$
    \EndIf
\EndFor

\State \Return $\varnothing$
\end{algorithmic}
\end{algorithm*}

\begin{table}[t]

\small
\setlength{\tabcolsep}{4pt}
\centering
\begin{tabular}{c c c c}
\hline
\textbf{Round} & \textbf{Relevant} & \textbf{Irrelevant} & \makecell{\textbf{Relevance}\\\textbf{Rate}} \\
\hline
1 & 674 & 179 & 79.02\% \\
2 & 456 & 109 & 80.71\% \\
3 & 448 & 117 & 79.29\% \\
4 & 487 & 139 & 77.80\% \\
5 & 472 & 154 & 75.40\% \\
\hline
\multicolumn{3}{l}{\textbf{Mean relevance rate}} & 78.44\% \\
\multicolumn{3}{l}{\textbf{Sample standard deviation (pp)}} & 1.99 \\
\multicolumn{3}{l}{\textbf{Coefficient of variation}} & 2.54\% \\
\hline
\end{tabular}
\caption{Verification results for taxonomy assignment. Summary statistics are computed across the five rounds; pp denotes percentage points.}
\label{tab:taxonomy_validation}
\end{table}

\subsection{Rule-assisted Candidate Screening}
\label{app:candidate_screening}
Rules are informed by the reviewer-selection experience of several editors of IEEE and ACM publications and applied automatically to publication records retrieved through the discipline taxonomy.
Editorial input concerns the screening procedure; it does not imply individual manual assessment of the full scholar pool.
Candidate groups are defined by the observable publication evidence described in Section~\ref{sec:task}.
Publication and indexing timelines in the source databases do not permit exact reconstruction of a candidate's publication count at every historical review date.

\section{Case Study for OmniReview}

\label{app:omni_example}
Table~\ref{tab:ofo_case_study} presents an example of peer review records from our OmniReview dataset. By leveraging identifiers from Open Academic Graph (OAG) and ORCID, we can access these databases to retrieve comprehensive metadata. Table~\ref{tab:oag_scholar_case_study} showcases an example of scholar metadata from Open Academic Graph, while Table~\ref{tab:oag_case_study} illustrates the paper metadata structure within the same database.

\section{Usage of the Released Dataset}
\label{app:dataset_usage}
Due to copyright constraints and to avoid duplicating large volumes of third-party data, we do not redistribute full author and paper metadata in the released benchmark. Instead, we provide OAG IDs and ORCID IDs pointing to the corresponding records in Open Academic Graph\footnote{\url{https://www.aminer.cn/open/article?id=5965cf249ed5db41ed4f52bf}} and ORCID\footnote{\url{https://info.orcid.org/documentation/integration-guide/working-with-bulk-data}}. We use OAG version 3.2. Each review record serves as a lightweight index: users obtain paper, reviewer, candidate, and editor identifiers from the benchmark, then use paper OAG IDs to retrieve article metadata, including titles, abstracts, venues, years, and references. Scholar OAG IDs link to researcher metadata and publication lists. Optional ORCID IDs support identity verification and profile completion. This design keeps the benchmark compact and avoids redistributing full third-party metadata. Appendix~\ref{app:omni_example} provides an example record; Tables~\ref{tab:oag_scholar_case_study} and~\ref{tab:oag_case_study} show example scholar and paper metadata.

\begin{table*}[t]
    \centering
    \small

    \setlength{\tabcolsep}{4pt}
    \renewcommand{\arraystretch}{1.08}
    \begin{tabularx}{\textwidth}{|>{\raggedright\arraybackslash}p{0.23\textwidth}|>{\raggedright\arraybackslash}X|}
        \hline
        Paper OAG ID & 643e5d140746dc40e37132a4\\
        \hline
        Author OAG IDs & 63b015c684ab04bd7fc3601a, 65c9ced7671a2719984cfa1d, 653291ef1ae766ad0180d649, 53f42ca2dabfaee02ac56175, 54409165dabfae805a6d2900\\
        \hline
        Author ORCIDs (optional) & 0000-0002-5962-4008, 0000-0002-6151-6858, 0000-0002-7268-9041, 0000-0002-4355-8368, 0000-0003-3100-1092\\
        \hline
        Reviewer OAG IDs & 53f4535fdabfaefedbb459b9, 62e495fcd9f204418d6ba64d\\
        \hline
        Reviewer ORCIDs (optional) & 0000-0002-1147-766X, 0000-0002-2845-8617\\
        \hline
        Unqualified Candidate OAG IDs & 54593219dabfaeb0fe344e7a, 640113cc45b36d5cbc41df62, 6403a6817691d561fb1f957b, 56319e4745ceb49c5e2136d8, 542d5076dabfae11fc45b850, 63afa47a84ab04bd7fb9a239, 64ca59ff75f2d36822961435, 670a2b79f4d6c71db9073468\\
        \hline
        Unqualified Candidate ORCIDs (optional) & None, 0000-0002-9899-7659, 0000-0003-1977-1545, 0000-0002-3807-3171, 0000-0003-2421-7290, None, 0000-0002-6742-1964, None\\
        \hline
        Potential Candidate OAG IDs & 64049d0b7691d561fb35e34f, 64902f4f58a6797656b4e24c, 53f3acd6dabfae4b34b0283b, 53f428c9dabfaeb1a7b607e0, 53f43a17dabfaeee229cd2d0, 53f488e4dabfaec09f2aec95, 53f44eb6dabfaec09f1e9159, 53f43daddabfaee2a1d1dd42\\
        \hline
        Potential Candidate ORCIDs (optional) & 0009-0001-6077-3924, 0000-0003-2442-3246, 0000-0002-4524-993X, 0000-0003-3689-8433, None, None, None, 0000-0001-5659-6841\\
        \hline
        Editor OAG ID & 53f43a11dabfaeecd6980dd1\\
        \hline
        Editor ORCID (optional) & None\\
        \hline
        L1 Category & Computer Science and Technology\\
        \hline
        L3 Category & Computer Neural Networks\\
        \hline
    \end{tabularx}

\caption{Example record from the OmniReview dataset showing peer review metadata with cross-referenced identifiers from Open Academic Graph (OAG) and ORCID. The record includes paper identification, author and reviewer information with their respective OAG IDs and ORCIDs (where available), candidate reviewer pools (unqualified and potential), editor details, and subject categorization at both L1 and L3 levels.}
\label{tab:ofo_case_study}
\end{table*}

\begin{table}[t]
    \centering
    \small

    \setlength{\tabcolsep}{4pt}
    \renewcommand{\arraystretch}{1.05}
    \begin{tabular}{|c|l|}
        \hline
        Scholar OAG ID & 53f4535fdabfaefedbb459b9 \\
        \hline
        Name & Maria A. Zuluaga \\
        \hline
        Org & EURECOM \\
        \hline
        ORCID & 0000-0002-1147-766X \\
        \hline
    \end{tabular}

\caption{Example scholar metadata from Open Academic Graph.}
\label{tab:oag_scholar_case_study}
\end{table}

\begin{table*}[t]
    \centering
    \small

    \setlength{\tabcolsep}{4pt}
    \renewcommand{\arraystretch}{1.08}
    \begin{tabularx}{\textwidth}{|>{\raggedright\arraybackslash}p{0.18\textwidth}|>{\raggedright\arraybackslash}X|}
        \hline
        Paper OAG ID & 643e5d140746dc40e37132a4\\
        \hline
        Paper Title & Multilevel comparison of deep learning models for function quantification in cardiovascular magnetic resonance: On the redundancy of architectural variations\\
        \hline
        Keywords & cardiovascular magnetic resonance, MRI, artificial intelligence, deep learning, cardiac image segmentation\\
        \hline
        Paper Abstract & Background Cardiac function quantification in cardiovascular magnetic resonance requires precise contouring of the heart chambers. This time-consuming task is increasingly being addressed by a plethora of ever more complex deep learning methods. However, only a small fraction of these have made their way from academia into clinical practice. In the quality assessment and control of medical artificial intelligence, the opaque reasoning and associated distinctive errors of neural networks meet an extraordinarily low tolerance for failure. Aim The aim of this study is a multilevel analysis and comparison of the performance of three popular convolutional neural network (CNN) models for cardiac function quantification. Methods U-Net, FCN, and MultiResUNet were trained for the segmentation of the left and right ventricles on short-axis cine images of 119 patients from clinical routine. The training pipeline and hyperparameters were kept constant to isolate the influence of network architecture. CNN performance was evaluated against expert segmentations for 29 test cases on contour level and in terms of quantitative clinical parameters. Multilevel analysis included breakdown of results by slice position, as well as visualization of segmentation deviations and linkage of volume differences to segmentation metrics via correlation plots for qualitative analysis. Results All models showed strong correlation to the expert with respect to quantitative clinical parameters (rz' = 0.978, 0.977, 0.978 for U-Net, FCN, MultiResUNet respectively). The MultiResUNet significantly underestimated ventricular volumes and left ventricular myocardial mass. Segmentation difficulties and failures clustered in basal and apical slices for all CNNs, with the largest volume differences in the basal slices (mean absolute error per slice: 4.2 $\pm$ 4.5 ml for basal, 0.9 $\pm$ 1.3 ml for midventricular, 0.9 $\pm$ 0.9 ml for apical slices). Results for the right ventricle had higher variance and more outliers compared to the left ventricle. Intraclass correlation for clinical parameters was excellent ($\geq 0.91$) among the CNNs. Conclusion Modifications to CNN architecture were not critical to the quality of error for our dataset. Despite good overall agreement with the expert, errors accumulated in basal and apical slices for all models.\\
        \hline
        Year & 2023\\
        \hline
        Venue & FRONTIERS IN CARDIOVASCULAR MEDICINE\\
        \hline
        Reference IDs & 617b66765244ab9dcbb6a772, 5c0f8b00da562944ac9c4db1, 58d82fcbd649053542fd5e6e, 5d1c7ccb3a55ac8c230aa97c, 5c04965717c44a2c74707951, 5c0f8cfdda562944aca09225, 622861f15aee126c0f853993, 627cd7a35aee126c0f3ff10f, 62652e325aee126c0fcfa6be, 5de0ec7e3a55acaf316bc74d, 59cb0d990cf23f44bec69750, 5fb255c691e01186d3f5df69, 62f6034590e50fcafdf6fc48, 62f23b8890e50fcafdca585c, 60ffe0c35244ab9dcb2cdc91, 612a13c15244ab9dcb1362f3, 5dca89823a55ac77dcb0213a, 5db92bcf47c8f7664623343a, 621865e35aee126c0fa5818e, 53e9a76db7602d97030a5fa6, 5c61603ce1cd8eae1501d8f9, 5fd0a271d4150a363cdde009, 60cdcb2de4510cd7c8655ca3, 5c6fc844e1cd8e35d214587d, 5550417d45ce0a409eb3bc08, 61dd5fae5244ab9dcb8bcd38, 5f48dde79fced0a24beafa07, 5c0f8893da562944ac975086, 573698016e3b12023e6da477, 61c301b85244ab9dcb04a914, 5f954e8e9fced0a24b179a94, 55a537a165ceb7cb02e4c52f, 5c136c7fda56295a08a8fead, 5dca89783a55ac77dcb01e84, 56d8e6c8dabfae2eee365661, 605aa2a4e4510cd7c86be281\\
        \hline
    \end{tabularx}

\caption{Example paper metadata from Open Academic Graph}
\label{tab:oag_case_study}
\end{table*}

\section{Details of Implemented Methods}
\label{app:baseline}
\textbf{Baselines.}
For TF-IDF, SciBERT, SPECTER2, BGE-M3, and Qwen3-Embedding-4B, we compute paper--reviewer scores in two steps. First, we calculate cosine similarity between the target paper representation and the candidate representation, obtained by averaging representations of the selected candidate's publications. Second, we fit isotonic regression on training pairs comprising historical reviewers and candidates who are semantically most distant within the discipline taxonomy. This maps similarity scores to confidence scores in $[0,1]$.

For CoF, we follow its original scoring pipeline and fit isotonic regression on positive and negative training pairs to map scores to $[0,1]$.

For RGCN, we initialize paper and scholar nodes with Qwen3-Embedding-4B representations. A DNN followed by softmax maps the graph's output embeddings to scores.

For LLM-as-Judge, we provide the LLM with the titles and abstracts of each candidate's five most recent publications to fit its context window because of the limited context window of LLM. We request a suitability score from 0 (no match) to 10 (a perfect match). Figure~\ref{fig:llm_judge_prompt} shows the prompt.

\textbf{Pro-MMoE Training Details.}
Our pipeline uses Qwen3-30B-A3B for paper summarization, Qwen3-Embedding-4B for semantic embeddings, and an MMoE module with three experts. For each candidate reviewer, we select up to ten publications using the five most highly cited and five most recent papers. In the Summary Branch, each selected publication is summarized and embedded independently, and the resulting publication embeddings are mean-pooled to represent the reviewer. The Raw-text Branch mean-pools embeddings of the same publications' original titles and abstracts. Figure~\ref{fig:paper_summary_prompt} shows the publication-level summarization prompt.

Training follows the two-stage strategy described with Equation~\ref{eq:training_objective}. Stage 1 uses the confidence objective for Tasks~1 and~2 to train the shared reranker modules and confidence gate, tower, and scorer. Stage 2 starts from the best Stage 1 checkpoint and optimizes the preference and qualification objectives for Tasks~3 and~4. The reported configuration freezes the shared encoders, feature builders, fusion module, experts, and confidence gate, tower, and scorer; only the two ranking gates, towers, and scorers are updated. The summarization and embedding models remain fixed, providing cached representations. Table~\ref{tab:dual_path_mmoe_hparams} lists the stage-specific loss weights and other hyperparameters.

\textbf{Hyperparameter Details.}
\label{app:hyperparameter}
We initialize all linear layers with Xavier uniform weights and zero biases. Hyperparameters are listed in Table~\ref{tab:dual_path_mmoe_hparams}.

\textbf{Compute Cost.}
All data processing, summary generation, embedding generation, and model inference were conducted on a server with four NVIDIA RTX A6000 GPUs.
Generating semantic embeddings for 110M papers using baseline methods, including BGE-M3 and Qwen3-Embedding-4B, took approximately 48 hours; generating CoF's multiple embeddings took approximately 96 hours.
Generating LLM summaries and semantic embeddings for 120K papers during Pro-MMoE preprocessing took approximately 30 hours.

With summaries and embeddings cached, scoring a single paper--reviewer pair takes approximately 100 ms.


\begin{table*}[p]
\begin{minipage}[t]{0.48\textwidth}
\vspace{0pt}
\centering
\small
\setlength{\tabcolsep}{5pt}

\begin{tabularx}{\columnwidth}{>{\raggedright\arraybackslash}X >{\raggedright\arraybackslash}p{0.34\columnwidth}}
\toprule
\textbf{Category} & \textbf{Value} \\
\midrule
Embedding dimension & 1024 \\
Reviewer encoder output dimension & 256 \\
Reviewer paper encoder & mean pooling\\
Maximum reviewer papers & 10 \\
Citation-based paper preselection top-$k$ & 5 \\
Recency-based paper preselection top-$k$ & 5 \\
Fusion MLP hidden dimensions & [512, 256] \\
Number of experts & 3 \\
Expert dimension & 128 \\
Expert MLP hidden dimensions & [512, 256] \\
Task tower hidden dimensions & [256, 128] \\
Dropout & 0.1 \\
Router noise & 0.05 \\
Stage 1 loss weights $(\lambda_0,\lambda_1,\lambda_2)$ & (1.0, 0.0, 0.0) \\
Stage 2 loss weights $(\lambda_0,\lambda_1,\lambda_2)$ & (0.0, 1.0, 1.0) \\
Hard-negative coefficient $\alpha$ & 2.0 \\
Hard-negative exponent $\gamma$ & 2.0 \\
Gate entropy weight & 0.01 \\
Gate load-balance weight & 0.1 \\
Optimizer & AdamW \\
Learning rate & $1\times10^{-4}$ \\
Weight decay & 0.01 \\
Learning-rate schedule & cosine decay with warmup \\
Warmup ratio & 0.1 \\
Gradient clipping & 1.0 \\
Batch size & 32 \\
Epochs & 40 \\
Stage 1 epochs & 20 \\
Stage 2 shared-module training & False \\
Random seed & 0 \\
\bottomrule
\end{tabularx}
\caption{Hyperparameter configuration for the dual-path paper-set MMoE reranker.}
\label{tab:dual_path_mmoe_hparams}
\end{minipage}
\hfill
\begin{minipage}[t]{0.48\textwidth}
\vspace{0pt}
\centering
\begin{lstlisting}[style=promptstyle]
SYSTEM_MESSAGE = """
You are an expert in academic paper review. Your task is to assess
whether a candidate is suitable for reviewing a target paper.
"""

USER_TEMPLATE = """
Target Paper:
Title: {}
Abstract: {}

The five representative papers of the candidate:
{}

Please provide a confidence score indicating the suitability of the
candidate for reviewing the target paper, based on the candidate's
research field, methods, and topic alignment with the target paper.

Output format:
Only output an integer between 0 and 10, where 10 indicates a perfect
match and 0 indicates no match at all. Do not output any other content.
"""
\end{lstlisting}
\captionof{figure}{Prompt for the LLM-as-Judge baseline.}
\label{fig:llm_judge_prompt}
\end{minipage}
\end{table*}

\begin{figure*}[t]
\centering
\begin{lstlisting}[style=promptstyle]
SYSTEM_MESSAGE = """
You are an expert academic reviewer assistant. Generate a concise,
critical summary of the research paper based ONLY on the title and
abstract provided. Your summary must help peer reviewers quickly assess
the paper's suitability for publication.

Critical requirements:
1. Strict word limit:
   Write 150--180 words. Every word must add value.

2. Source restriction:
   Use ONLY the provided title and abstract. Never invent details,
   methods, results, or context not explicitly stated.

3. Reviewer-focused content:
   - Core research question and significance
   - Methodology assessment based on the abstract
   - Key findings and their reliability
   - Novel contributions relative to existing work
   - Major limitations explicitly mentioned
   - Overall publication recommendation likelihood

4. Tone and style:
   - Objective and professional academic tone
   - Critical but constructive evaluation
   - Avoid promotional language or excessive praise
   - Use precise terminology from the field

5. Structure:
   Write a single coherent paragraph following the order:
   Problem -> Methods -> Findings -> Contribution -> Limitations -> Assessment.

Warning:
If the abstract lacks crucial details, such as methods, limitations, or
contributions, explicitly state these gaps as review concerns. Never
compensate with external knowledge.
"""

USER_TEMPLATE = """
Please provide a critical review summary for peer reviewers based on the
following paper title and abstract. Strictly adhere to the word limit and
guidelines.

Title: "{}"
Abstract: "{}"
"""
\end{lstlisting}
\caption{Prompt for Paper Summarization.}
\label{fig:paper_summary_prompt}
\end{figure*}

\end{document}